\documentclass[pra,aps,preprint]{revtex4-1}
\usepackage[T1]{fontenc}
\usepackage[latin9]{inputenc}
\usepackage[active]{srcltx}
\usepackage{amsmath}
\usepackage{amssymb}
\usepackage{esint}
\usepackage{lineno}
\usepackage{color,soul}
\usepackage{xcolor}
\makeatletter
\usepackage{epsf,epsfig}
\usepackage{graphicx}

\makeatother
\begin{document}
\title{Influence of non-linearity of medium on the laser induced filamentation instability in magnetized plasma}
\author{Sepideh Dashtestani, Akbar Parvazian}
\affiliation{Department of  Physics, Isfahan University of Technology, Isfahan 84156-83111, Iran}\author{Hamidreza Mohammadi} \thanks{hr.mohammadi@sci.ui.ac.ir} \affiliation{Department of Physics, University of Isfahan, Isfahan, Iran} \affiliation{Quantum Optics Group, University of Isfahan, Isfahan, Iran}
\date{\today}
\begin{abstract}
\noindent The effects of the non-linearity of the medium on the growth rate of filamentation instability in a magnetized plasma interacting with an intense laser pulse, is investigated. The non-linearity of the medium, modeled by Kerr non-linearity, is an important factor, which determines the rate of instability growth. Sensitivity of the rate of filamentation growth, to the Kerr non-linear coefficient could be adjusted by the external magnetic field and laser intensity.
\end{abstract}
\maketitle
\section{Introduction}
Laser-induced instabilities in plasma is a vast studied subject in the field of laser-plasma interaction. Among these instabilities, filamentation, a large-scaled phenomenon propagation along the direction of the laser light, is one of the most important. First time the  filamentation and spectral (red) broadening phenomena were reported by  Shimizu \textit{et. al}\cite{Ref1}. Authors of Ref. \cite{Ref2} have been showed that a powerful femtosecond laser pulse could uncover a sub-field of applied physics, namely, filamentation non-linear optics. Controlling this phenomenon (filamentation) can be used 
to improve the performance of the laser lightning protection devices \cite{Ref3}, self channeling and selfguiding laser system  \cite{Ref4,Ref5} and etc.
The filamentation in various media such as air, water and 
gases have been heavily studied in recent years  \cite{Ref6,Ref7}. The filamentation phenomenon originates from competition between self-focusing and plasma defocusing of an intense laser propagating in Kerr medium \cite{Ref75}.
In this paper we investigate formation of the filamentation induced by
laser in magnetized plasma, by relating  the instability 
growth rate of the filamentation to the controllable parameters of the
medium, the laser and also the external magnetic field.
In a laser plasma interaction the presence of the magnetic 
field was seen to impress the filamentation very remarkably 
\cite{Ref8,Ref9}. The magnetic field can influence 
the transverse size and growth rate of filaments.
The output results of 30 years investigation in this field reveal that the laser induced multiple filamentation initialization stems to random noise existing in the input beam profile \cite{Ref10}. Despite this explanation, some works states that multiple filamentation was created 
due to vectorial effects \cite{Ref11,Ref12,Ref13}. Numerical simulations show that when the input 
beam is sufficiently powerful, vectorial effects lead to multiple filamentation. According to the results of this works, non-linear response of the medium to the laser pulse, appears to be a strong determinant of multiple filamentation. The effects of the non-linearity of the magnetized plasma on the formation and the growth of filamentation are not considered yet. In this paper we study the growth rate  of the laser induced filamentation in a magnetized plasma and bold the effects of non-linearity of the medium on the filamentation formation. In this way we compare 
the growth rate of the filamentation in linear and non-linear mediums.
The system under consideration is a plasma which is deriving by
an external magnetic field, directed in +z-direction and interacted by an
intense pulsed laser light propagates in the same direction.
The laser field assumed to has right-circular polarization (RCP).
The results show that beside the noise effects the non-linear 
response of the medium has a significance effects on the
construction and growth rate of the filamentation.
We assume that plasma medium is influenced by a magnetic field
$\vec{B}=(0,0,B) $ and parametrized by: 
$\omega_p=(\frac{4 \pi n_0e^2}{m})^\frac{1}{2}$ and 
$\omega_c=\frac{eB}{mc}$ as plasma frequency and cyclotron frequency
of plasma with electron density $n_0$, respectively.
Here, -e, m are the charge and mass of the electron, respectively.
We consider a circularly polarized laser beam propagated in the direction of
external magnetic field:
$\vec{E_0} =( \hat{x} +i \hat{y})A_{10}e^{-i( kz-\omega t)}$
where, $k=\frac{\omega }{c}[1-\frac{\omega _{p}^2}{\omega ^2
(1-\frac{\omega _c}{\omega })}]^\frac{1}{2}$.
Where $A_{01}$ is a real constant proportional to square root of laser intensity.
We amuse that the laser intensity has an instability modeled by
considering a circularly polarized ripple in the
electric field amplitude,
$\vec{E}=\vec{E_0}+\vec{E_1}$.
Where, $\vec{E_1}=( \hat{x} +i \hat{y})A_{1}(x,z)e^{-i( \omega t-kz)}$
and $A_{10}$ is a complex number.
The laser pulse imparts the electron to oscillate with velocity:
\begin{equation} \label{eq1}
\vec{v}=\frac{e\vec{E}}{mi(\omega -\omega _c)}.
\end{equation}
The pondermotive force is \cite{Ref14}:
\begin{equation}
\vec{f}_p=-e\frac{\vec{v}\times \vec{B}}{c}-m(\vec{v}.\vec{\bigtriangledown})\vec{v}.
\end{equation}
The time independent part of $\vec{f}_p$ can be written as
\begin{equation}
\vec{f}_p=\frac{-e^2}{4m(\omega-\omega_c)^2}\vec{\bigtriangledown}(\vec{E}\vec{E}^*),
\end{equation}
where $*$ represents complex conjugate. Using Eq. (\ref{eq1}) 
for velocity $\vec{v}$ and 
$\vec{B}=\frac{\vec{\bigtriangledown}\times \vec{E}}{i\omega}$, 
the pondermotive force is given by
\begin{equation}
\vec{f}_p=\frac{-e^2}{4m(\omega-\omega_c)^2}
\vec{\bigtriangledown}A_{10}(A_1+A^*_1),
\end{equation}
and the corresponding pondermotive potential is
\begin{equation}
\phi _p=\frac{-e}{4m(\omega-\omega_c)^2}A_{10}(A_1+A^*_1).
\end{equation}
The electrons was expelled away from the regions of 
higher electric field by this pondermotive 
potential while ions remain stationary due to 
their large inertia \cite{Ref14}. An electrostatic 
field $\vec{E}=-\vec{\bigtriangledown}\phi$, which is 
induced by space charge, pulls back the electrons
causes plasma oscillation. 
Under the quasisteady state condition and cold 
plasma approximation we can following  \cite{Ref14} 
and write $\phi=-\phi_p$ . The modified electron 
density could be calculated from the Poisson's equation, 
$\bigtriangledown^2\phi=4 \pi e(n_e-n_0)$, is:
\begin{equation}
n_e=n_0-\frac{1}{4\pi e}\bigtriangledown^2\phi_p.
\end{equation}
In the magnetized plasma the effective dielectric 
constant for the right circularly polarized 
extraordinary mode by use the Lorentz-Durde model can be written as \cite{Ref15}:
\begin{equation}
 \varepsilon _{+}=1-\frac{n_e\omega^2_p}{n_0\omega^2(1-\frac{\omega_c}{\omega})}.
\end{equation} 
Substituting for $n_e$ from Eq. (5) in Eq. (6) , now we have
\begin{equation}
 \varepsilon _{+}= \varepsilon _{0+}+\psi _{+}(A_{10}(A_1+A_1^*))
\end{equation}
Now we can consider linear and non-linear part
of the dielectric constant as follows \cite{Ref14}:
\begin{eqnarray}
\varepsilon _{0+}&=&1-\frac{\omega _{p}^2}{\omega ^2(1-\frac{\omega _c}{\omega })},\nonumber
\\
\psi _{+}(A_{10}(A_1+A_1^*))&=&
-\frac{e^2\vec{\triangledown }^2A_{10}(A_{1}+A_{1}^*)}
{4m^2\omega ^5(1-\frac{\omega _c}{\omega})^3}.\nonumber
\end{eqnarray}
Where $\psi _{+}(A_{10}(A_1+A_1^*))$ is the non-linearity
induced by electric field ripple 
of the dielectric constant and $  \varepsilon _{0+}$ is 
the linear response of the medium.
The response of the medium to the laser light is encapsulated
in dielectric constant $  \varepsilon $ or electric susceptibility $\chi $.
The electric constant can be calculated by the concept of pondermotive 
force and Maxwell equations.
propagation of an electromagnetic wave in a dielectric medium is governed 
by the vector wave equation extracted from  Maxwell equations.
The vector wave equation for $\overrightarrow{E}$
in a homogeneous medium can be expressed as \cite{Ref16,Ref17}:
\begin{equation} \label{wave equation}
\triangledown^{2}\vec{E}-\vec{\triangledown }
(\vec{\triangledown}\cdot \vec{E})=\frac{1}{c^2}
\frac{\partial ^2 \vec{E}}{\partial t^2}+\frac{4 \pi }{c^2}
\frac{\partial^2 \vec{P}}{\partial t^2}+\frac{4 \pi }{c^2}
\frac{\partial \vec{J}}{\partial t},
\end{equation}
where $\vec{P}=\vec{P}^{(1)}+\vec{P}_{NL}$ 
and $\vec{j}=-n_ee\vec{v}$.
\begin{equation}\label{wave equation1}
\triangledown^{2}\vec{E}-\vec{\triangledown }
(\vec{\triangledown}\cdot \vec{E})=\frac{1}{c^2}
\frac{\partial ^2 \vec{E}}{\partial t^2}+ \frac{4 \pi }{c^2}
\frac{\partial ^2\vec{P} \,^{(1)}}{\partial t^2}+
\frac{4 \pi }{c^2}\frac{\partial \vec{J}}{\partial t}+
\frac{4 \pi }{c^2}\frac{\partial ^2\vec{P}_{NL}}{\partial t^2}.
\end{equation}
In the following, in section 2, the filamentation 
formation rate in a linear medium ($\vec{P}_{NL}=0$) in the presence of laser noise is calculated. Then, the effects of 
non-linearity of the medium are illustrated in section 3.
Ultimately, the relation between the growth rate of the 
filamentation and the system and environment parameters is presented in section 4.
Finally, in section 5, a discussion concludes the paper.

\section{linear Response of the medium}

\begin{figure}[!ht]
\centering
\includegraphics[width=9cm]{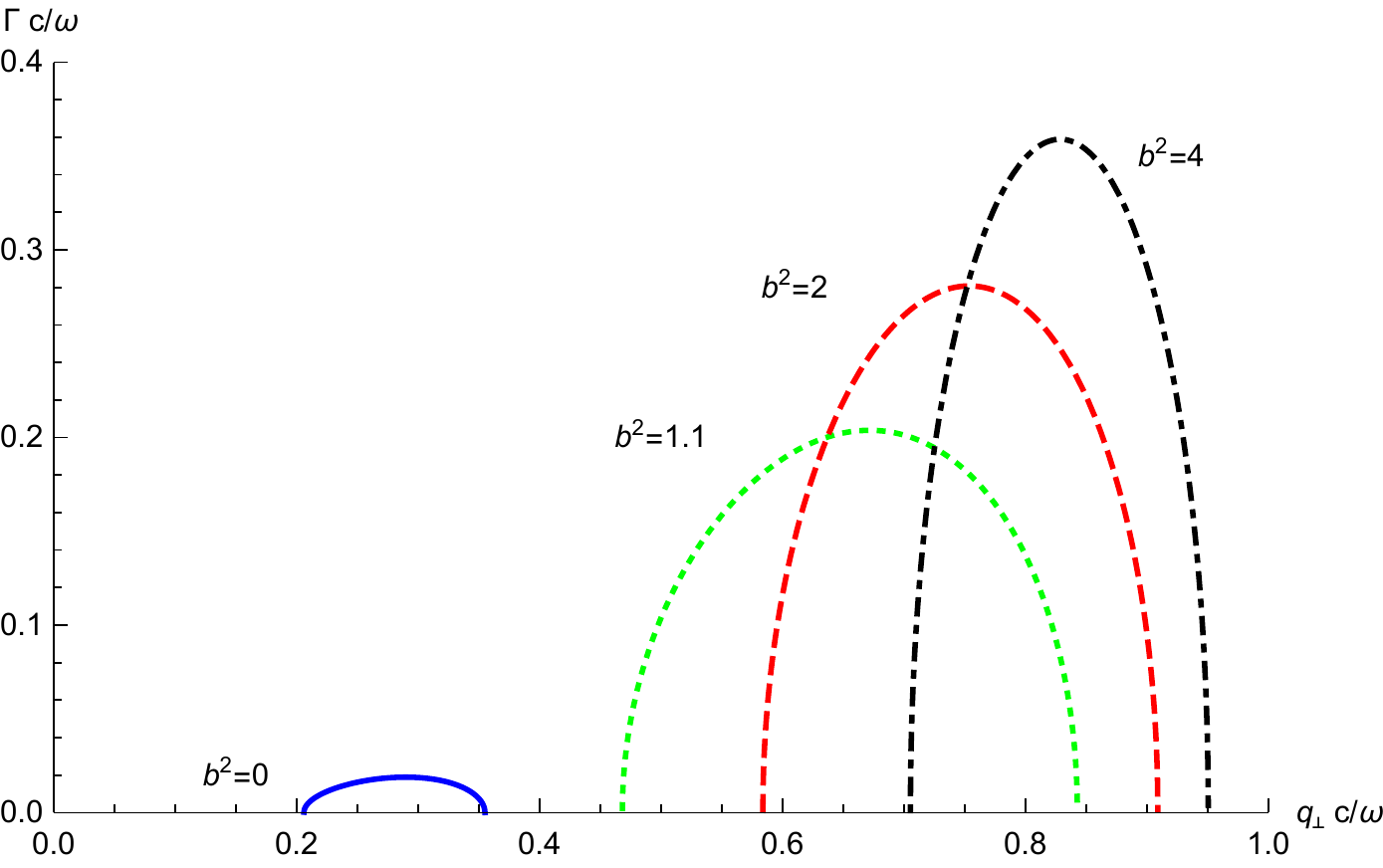}
\caption{ (Color online) normalized growth 
rate $\Gamma _{nor}(\Gamma c/\omega)$ versus 
normalized transverse wave number $Q_L(q_\perp  c/\omega)$
for $\omega _{p}^2/ \omega ^2=0.1$, $\omega _c / \omega=0.2, a_0=1$
, $d=\frac{1}{2}$ and $\gamma=\frac{1}{2}$.}
\label{fig1}
\end{figure}

In the low-intensity regime the response of the medium (magnetized plasma)
is linear, i.e. the refractive index is intensity-independent.
Thus for linear medium we have $\vec{P}_{NL}=0$
and we can write\cite{Ref18}:  
\begin{equation}
\vec{p} \,^{(1)} = \frac{1}{4 \pi} \,\chi ^{(1)} \vec{E},
\end{equation}
where the constant of proportionality $\chi ^{(1)}$, is known 
as the linear susceptibility, $ \varepsilon_0$ is the permittivity 
of Vacuum and $n_0$ is the linear refractive index of medium. 
By considering $1+\chi ^{(1)}= \varepsilon$ from Eq. (\ref{wave equation}) we have:
\begin{equation}
\triangledown^{2}\vec{E}-\vec{\triangledown }
(\vec{\triangledown}\cdot \vec{E})+\frac{\omega ^2}
{c^2}\epsilon \cdot\vec{E}-\frac{4 \pi }{c^2}\frac{\partial \vec{J}}
{\partial t}=0.
\end{equation}
By using $ \frac{\partial E_z}{\partial Z} =
-( \frac{1}{ \varepsilon _{zz}})[( \frac{\partial}{\partial x} )
( \varepsilon _{xx}E_x+ \varepsilon_{xy}E_y)+
( \frac{\partial}{\partial y} ) \times (- \varepsilon _{xy}E_x+
 \varepsilon _{xx}E_y)] $ in \cite{Ref14},
we obtain the following differential
equation for the field amplitude,
$A_+=(A_{10}+A_1)exp^{-i(kz-\omega t)}$:
\begin{equation}
\frac{ \partial ^2A_+}{ \partial z^2}+\frac{1}{2}
(1+\frac{ \varepsilon_{0+}}{ \varepsilon_{zz}})\triangledown_{\perp}^2A_+
+ \frac{\omega ^2}{c^2}(\psi _{+}A_{+})+\frac{\omega ^2}{c^2}
\frac{\omega ^2_p}{\omega ^2}\frac{d}{(1-\frac{\omega _c}{\omega})}A_+=0.
\end{equation} 
Here, subscript $\perp$ stands for traverse to the z-direction. Finally, the wave equation for perturbed field given by:
\begin{equation}
2ik\frac{\partial A_1}{\partial z}+\frac{1}{2}
(1+\frac{ \varepsilon_{0+}}{\epsilon_{zz}})\triangledown_{\perp}^2A_1+
\frac{\omega ^2}{c^2}(\psi _{+}A_{10})+
\frac{\omega ^2}{c^2}\frac{\omega ^2_p}
{\omega ^2}\frac{d}{(1-\frac{\omega _c}{\omega})}A_1=0.
\end{equation} \label{NLSE}
This is the famous Non-Linear Schr\"odinger Equation(NLSE). For the purpose of calculating of the growth rate of the filament, we separate the real and imaginary parts of Eq. (\ref{NLSE}).
By defining $A_1=A_{1r}+A_{1i}$, we have:
\begin{equation}
-2k\frac{\partial A_{1i}}{\partial z}+
\frac{1}{2}(1+\frac{ \varepsilon_{0+}}{ \varepsilon_{zz}})
\triangledown_{\perp}^2A_{1r}-\frac{e^2 \triangledown^2A_{10}^2(A_{1r}+
A_{1r}^*)}{4m^2\omega ^2c^2(1-\frac{\omega _c}{\omega})^3}+\frac{\omega ^2}
{c^2}\frac{\omega ^2_p}{\omega ^2}\frac{d}{(1-\frac{\omega _c}{\omega})}A_{1r}=0,\nonumber
\end{equation}
\begin{equation}
2k\frac{\partial A_{1r}}{\partial z}+\frac{1}{2}
(1+\frac{ \varepsilon_{0+}}{ \varepsilon_{zz}})\triangledown_{\perp}^2 A_{1i}+
\frac{\omega ^2}{c^2}\frac{\omega ^2_p}{\omega ^2}\frac{d}
{(1-\frac{\omega _c}{\omega})}A_{1i}=0, \nonumber
\end{equation}
where $A_{1i},A_{1r} \approx exp[i(q_\parallel z+q_\perp x)]$.
By replacing $\triangledown_\perp ^2 \rightarrow$ $-q_\perp ^2$
and $\frac{\partial }{\partial z} \rightarrow$ $iq_{\parallel }$ 
in the above equations the following coupled equations
for $A_{1r}$, and $A_{1i}$ has been achieved:
\begin{equation}
-2ikq_\parallel A_{1i}-\frac{1}{2}(1+\frac{ \varepsilon_{0+}}
{ \varepsilon_{zz}})q_\perp^2A_{1r}+\frac{a^2A_{1r}
(q_\perp^2+q_\parallel^2)}{2(1-\omega _c/\omega)}+
\frac{\omega ^2}{c^2}\frac{\omega ^2_p}{\omega ^2}
\frac{d}{(1-\frac{\omega _c}{\omega})}A_{1r}=0, \nonumber
\end{equation}
\begin{equation}
2ikq_\parallel A_{1r}-\frac{1}{2}(1+\frac{ \varepsilon_{0+}}
{ \varepsilon_{zz}})q_\perp^2A_{1i}+\frac{\omega ^2}{c^2}
\frac{\omega ^2_p}{\omega ^2}\frac{d}{(1-\frac{\omega _c}{\omega})}A_{1i}=0.\nonumber
\end{equation}
Where $a_0=\frac{eA_{10}}{m\omega c}$  is 
the normalized laser field amplitude,
$a^2=\frac{a^2_0}{(1-\omega _c/\omega)^2}$
and $d=\frac{n_e}{n_0}$.
\begin{eqnarray}
&[&\frac{4k^2}{1/2(1+ \varepsilon_{0+}/ \varepsilon_{zz})q_\perp^2-
\frac{\omega ^2}{c^2}\frac{\omega ^2_p}{\omega ^2}
\frac{d}{(1-\frac{\omega _c}{\omega})}}+\frac{a^2}
{2(1-\omega _c/\omega)}]q_\parallel^2 \nonumber\\
&+&[\frac{a^2}{2(1-\omega _c/\omega)}-
1/2(1+ \varepsilon_{0+}/ \varepsilon_{zz})]q_\perp^2+
\frac{\omega ^2}{c^2} \frac{\omega ^2_p}{\omega ^2}
\frac{d}{(1-\frac{\omega _c}{\omega})}=0.\nonumber
\end{eqnarray}
These equations gives the spatial growth rate $\Gamma$ as,
\begin{eqnarray}
\Gamma &=&iq_\parallel=[(\frac{a^2}{2(1-\omega _c/\omega)}
-1/2(1+ \varepsilon_{0+}/ \varepsilon_{zz}))q_\perp^2+
\frac{\omega ^2}{c^2}\frac{\omega ^2_p}{\omega ^2}
\frac{d}{(1-\frac{\omega _c}{\omega})}]^{\frac{1}{2}}\nonumber\\
&\times&[\frac{(1-\omega _c/\omega)(1+ \varepsilon_{0+}
/ \varepsilon_{zz})q_\perp^2-2\omega ^2/c^2\omega ^2_p/
\omega ^2d}{8k^2(1-\omega _c/\omega)+1/2a^2(1+ \varepsilon_{0+}
/ \varepsilon_{zz})q_\perp^2-\omega ^2/c^2(\omega ^2_p/
\omega ^2)da^2/(1-\omega _c/\omega)}]^{\frac{1}{2}}.
\end{eqnarray}
Solid lines in Figure \ref{fig1} illustrate the normalized growth rate
$\Gamma _{nor}(\Gamma c/\omega)$ versus
normalized transverse wave number $ Q_L(q_\perp  c/\omega)$
for some special fixed parameters.
\section{non-linear Response of the medium}
Where the laser power exceeds the critical power $P_c=\frac{3.77 \lambda^2}{8 \pi n_0 n_2}$
\cite{Ref19} the non-linear effects appears due to self-focusing phenomenon .
In this regime the refractive index of the medium depends on the laser
intensity and kerr phenomenon occurs.
This critical laser power can be easily achieved by the
ultra short femtosecond lasers.
The response of our system, magnetic plasma,
to such intense laser is governed by Eq. (10)
with $\vec{P}_{NL} \neq 0$.
\begin{equation}
\triangledown^{2}\vec{E}-\vec{\triangledown }
(\vec{\triangledown}\cdot \vec{E})+\frac{\omega ^2}{c^2}
 \varepsilon \vec{E}+\frac{4 \pi }{c^2}\omega ^2 \vec{P}_{NL}-\frac{4 \pi }{c^2}\frac{\partial \vec{J}}{\partial t}=0,
\end{equation} 
For isotropic and homogeneous Kerr medium we have \cite{Ref20,Ref21}:
\begin{equation}
\vec{P}_{NL}=\frac{4 n_0n_2}{4 \pi(1+\gamma)}
[(\vec{E}\cdot \vec{E}^{*})\vec{E}+
\gamma (\vec{E}\cdot \vec{E})\vec{E}^{*}],
\end{equation} 
where $n_2$ is the Kerr coefficient, $\vec{E}^{*}$ 
is the complex conjugate of $ \vec{E}$ and $\gamma$ 
is a positive constant whose value depends on the 
physical origin of the Kerr effect \cite{Ref11}.
\begin{equation}
\vec{\triangledown }\cdot \vec{E}=-\frac{4 \pi}
{n_0^2 }\vec{\triangledown }\cdot \vec{P}_{NL}
-4 \pi  \vec{ \bigtriangledown }.\vec{P} \,^{(1)}.
\end{equation}
The wave equation can be written as
\begin{eqnarray}
&\triangledown^2\vec{E}&+-\vec{\triangledown }
(\vec{\triangledown}\cdot \vec{E})+
\frac{\omega ^2}{c^2} \varepsilon \vec{E}\nonumber\\
&+&\frac{\omega^2}{c^2} \frac{4 n_0n_2}{(1+\gamma)}[(\vec{E}\cdot 
\vec{E}^{*})\vec{E}+\gamma (\vec{E}\cdot \vec{E})\vec{E}^{*}]+i\omega\frac{4 \pi }{c^2}\vec{J}=0,
\end{eqnarray} 
\begin{eqnarray}
&\frac{ \partial ^2A_+}{ \partial z^2}&+\frac{1}{2}
(1+2\frac{\varepsilon_{0+}}{\varepsilon_{zz}})
\triangledown_{\perp}^2A_+
+\frac{\omega ^2}{c^2}\varepsilon_0A_+
+\frac{\omega ^2}{c^2}\frac{4n_0n_2}{1+\gamma}
[(A_+\cdot A_+^*)A_+ +\gamma (A_+\cdot A_+)A_+^*]\nonumber\\
&+&\frac{\omega ^2}{c^2}\frac{\omega_p ^2}
{\omega^2}\frac{d}{(1-\omega_c/\omega)}A_{+}
=\frac{-4n_2}{n_0(1+\gamma )}[\vec{\triangledown }
(\vec{\triangledown }\cdot 
(A_+\cdot A_+^*)A_++\gamma (A_+\cdot A_+)A_+^*)].
\end{eqnarray}
The wave equation for perturbed field is given by
\begin{eqnarray}
&2ik&\frac{\partial A_1}{\partial z}+
\frac{1}{2}(1+2\frac{\varepsilon_{0+}}{\varepsilon_{zz}})
\triangledown_{\perp} ^2A_1+\frac{\omega ^2}{c^2}
(\psi _{+}A_{10})+\frac{\omega ^2}{c^2}\frac{4n_0n_2}{1+\gamma }[(A_{10})^2A_1
+\gamma (A_{10})^2A_1^*]\nonumber\\
&+&\frac{\omega ^2}{c^2}\frac{\omega ^2_p}
{\omega ^2}\frac{d}{(1-\frac{\omega _c}{\omega})}A_1
=\frac{-4n_2}{n_0(1+\gamma )}[\vec{\triangledown }
(\vec{\triangledown }\cdot((A_{10})^2A_1+\gamma(A_{10})^2A_1^*))].
\end{eqnarray} 
Separating the real and imaginary part we have:
\begin{eqnarray}
&-2k&\frac{\partial A_{1i}}{\partial z}
-\frac{1}{2}(1+2\frac{\varepsilon_{0+}}{\varepsilon_{zz}})
q_{\perp }^2A_{1r}
+\frac{a^2}{2(1-\omega _c/\omega)}q_\perp ^2A_{1r}+
\frac{a^2}{2(1-\omega _c/\omega)}q_\parallel ^2A_{1r} \nonumber\\
&+&\frac{\omega ^2}{c^2} \frac{4n_2n_0}{(1+\gamma)}
[(A_{10})^2 +\gamma (A_{10})^2]A_{1r}+\frac{\omega ^2}{c^2}\frac{\omega_p ^2}{\omega^2}\frac{d}{(1-\omega_c/\omega)}A_{1r}\nonumber \\
&+&\frac{4n_2}{n_0(1+\gamma)}[(A_{10})^2 \triangledown _\perp ^2+\gamma (A_{10})^2\triangledown _\perp ^2]A_{1r}=0,
\end{eqnarray}
\begin{eqnarray}
&2k&\frac{\partial A_{1r}}{\partial z}
-\frac{1}{2}(1+2\frac{\varepsilon_{0+}}{\varepsilon_{zz}})
q_{\perp }^2A_{1i}
+\frac{\omega ^2}{c^2}\frac{4n_0n_2}{1+\gamma }
[(A_{10})^2  -\gamma (A_{10})^2]A_{1i}
\nonumber\\
&+&\frac{\omega ^2}{c^2}\frac{\omega_p ^2}
{\omega^2}\frac{d}{(1-\omega_c/\omega)}A_{1i}+\frac{4n_2}{n_0(1+\gamma )}[(A_{10})^2\triangledown _\perp^2-\gamma (A_{10})^2\triangledown _\perp^2]A_{1i}=0,
\end{eqnarray}
\begin{eqnarray}
&-2ikq_\parallel& A_{1i}
-\frac{1}{2}(1+2\frac{\varepsilon_{0+}}{\varepsilon_{zz}})
q_{\perp }^2A_{1r}
+\frac{a^2}{2(1-\omega _c/\omega)}q_\perp ^2A_{1r}+\frac{a^2}{2(1-\omega _c/\omega)}q_\parallel ^2A_{1r}\nonumber\\
&+&\frac{\omega ^2}{c^2}\frac{\omega_p ^2}{\omega^2}\frac{d}{(1-\omega_c/\omega)}A_{1r}+\frac{\omega ^2}{c^2}4n_0n_2(A_{10})^2A_{1r}-\frac{4n_2}{n_0}(A_{10})^2q_\perp ^2A_{1r}=0,
\end{eqnarray}
\begin{equation}
2ikq_\parallel A_{1r}
-\frac{1}{2}(1+2\frac{\varepsilon_{0+}}{\varepsilon_{zz}})
q_{\perp }^2A_{1i}
+\frac{\omega ^2}{c^2}
\frac{4n_0n_2}{3}(A_{10})^2A_{1i}-
\frac{4n_2}{3n_0}(A_{10})^2q_\perp ^2A_{1i}
+\frac{\omega ^2}{c^2}\frac{\omega_p ^2}{\omega^2}\frac{d}{(1-\omega_c/\omega)}A_{1i}=0.
\end{equation} 
In the following, we define $b^{2}={4n_2}(A_{10})^2$ for the purpose of simplification.
\begin{eqnarray}
&[&\frac{4k^2}{\frac{b^2}{3n_0}q_\perp^2
+\frac{1}{2}(1+2\frac{\varepsilon_{0+}}{\varepsilon_{zz}})
q_{\perp }^2
-\frac{\omega ^2b^2n_0}{3c^2}-
\frac{\omega ^2}{c^2}\frac{\omega_p ^2}{\omega^2}\frac{d}
{(1-\omega _c/\omega)}}+\frac{a^2}{2(1-\omega_c/\omega)}]q_\parallel^2\nonumber\\
&+&[\frac{\omega ^2}{c^2}[\frac{\omega_p ^2}
{\omega ^2}\frac{d}{(1-\omega_c/\omega)}+n_0b^2]
+[-\frac{1}{2}(1+2\frac{\varepsilon_{0+}}{\varepsilon_{zz}})
+\frac{a^2}{2(1-\omega_c/\omega)}-
\frac{b^2}{n_0}]q_\perp ^2=0.
\end{eqnarray}
This equation gives the spatial growth rate $\Gamma$ as: 
\begin{eqnarray}
\Gamma &=&iq_\parallel =[\frac{\omega ^2}{c^2}
(\frac{\omega_p ^2}{\omega ^2}
\frac{d}{(1-\omega_c/\omega)}+b^2)+
(\frac{a^2}{2(1-\omega_c/\omega)}-b^2
-\frac{1}{2}(1+2\frac{\varepsilon_{0+}}{\varepsilon_{zz}}))q_\perp^2]^{\frac{1}{2}}\nonumber\\
&\times&\frac{[(1+2\frac{\varepsilon_{0+}}
{\varepsilon_{zz}})(1-\omega_c/\omega)+
(1-\omega_c/\omega)\frac{2b^2}{3}]q_\perp ^2-
\frac{\omega ^2}{c^2}(1-\omega_c/\omega)
\frac{2b^2}{3}-\frac{\omega ^2}{c^2}
\frac{\omega_p ^2}{\omega ^2}2d]^{\frac{1}{2}}}{[8k^2(1-\omega_c/\omega)+
[\frac{1}{2}(1+2\frac{\varepsilon_{0+}}{\varepsilon_{zz}})a^2
+\frac{a^2b^2}{3}]q_\perp ^2-\frac{\omega ^2}{c^2}
(\frac{a^2b^2}{3}-\frac{a^2d}{(1-\omega_c/\omega)}
\frac{\omega_p ^2}{\omega ^2})]^{\frac{1}{2}}}.
\end{eqnarray}

\section{Results}

\begin{figure}[!ht]
\centering
\includegraphics[width=17cm]{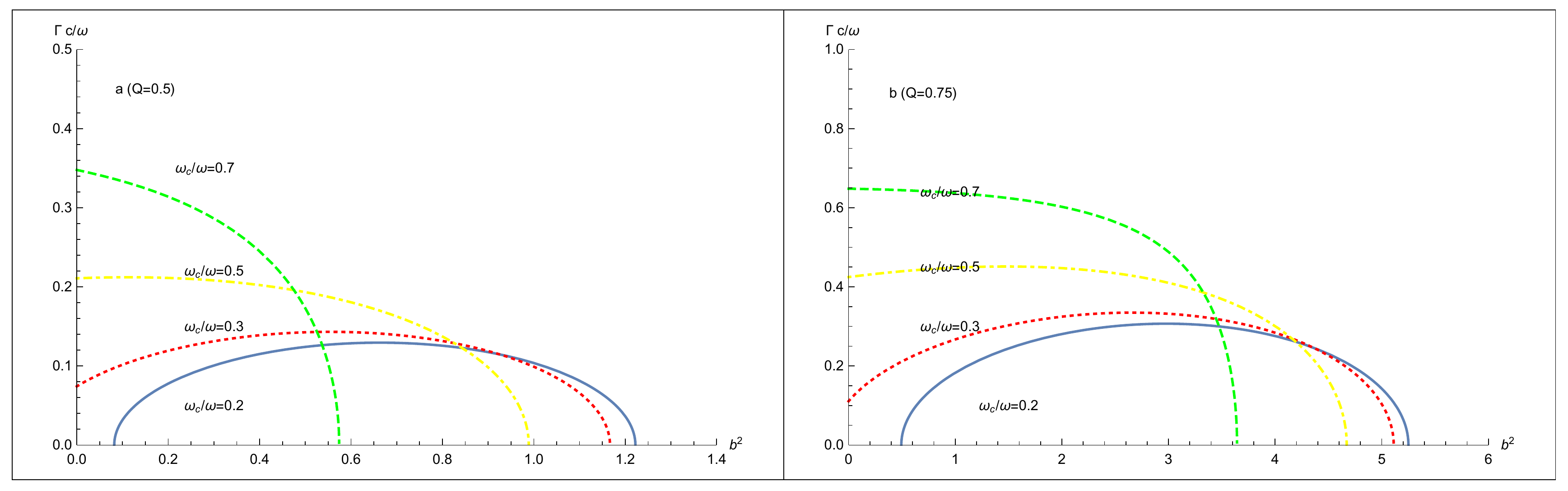}
\hspace{2mm}
\caption{ (Color online) normalized growth rate 
$\Gamma _{nor}(\Gamma c/\omega)$ versus $b^2$
for $\omega _{p}^2/ \omega ^2=0.1$, $a_0=1$, $d=\frac{1}{2}$, 
$Q_{ \perp }$= (2a :0.5, 2b :0.75) and $\gamma=\frac{1}{2}$.}
\label{fig2}
\end{figure}

\begin{figure}[!ht]
\centering
\includegraphics[width=17cm]{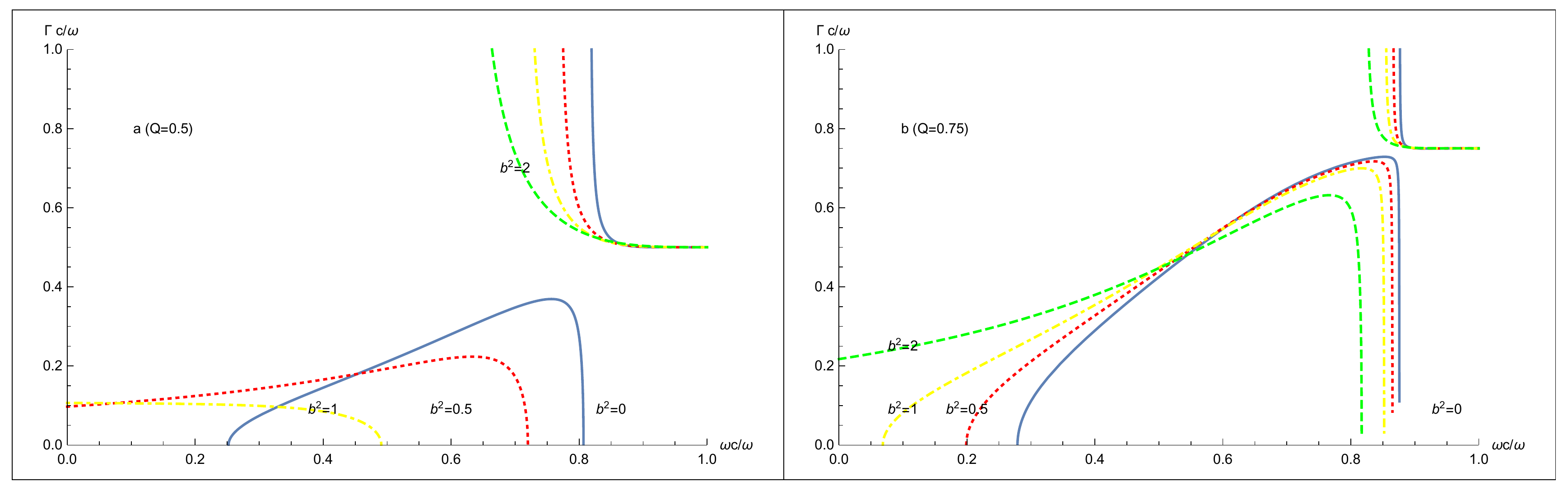}
\hspace{2mm}
\caption{ (Color online) normalized growth rate 
$\Gamma _{nor}(\Gamma c/\omega)$ versus $\frac{\omega_c}{\omega}$
for $\omega _{p}^2/ \omega ^2=0.1$, $a_0=1$, 
$d=\frac{1}{2}$, $Q_{ \perp }$=(3a :0.5, 3b :0.75) and $\gamma=\frac{1}{2}$.}
\label{fig3}
\end{figure}

\begin{figure}[!ht]
\centering
\includegraphics[width=17cm]{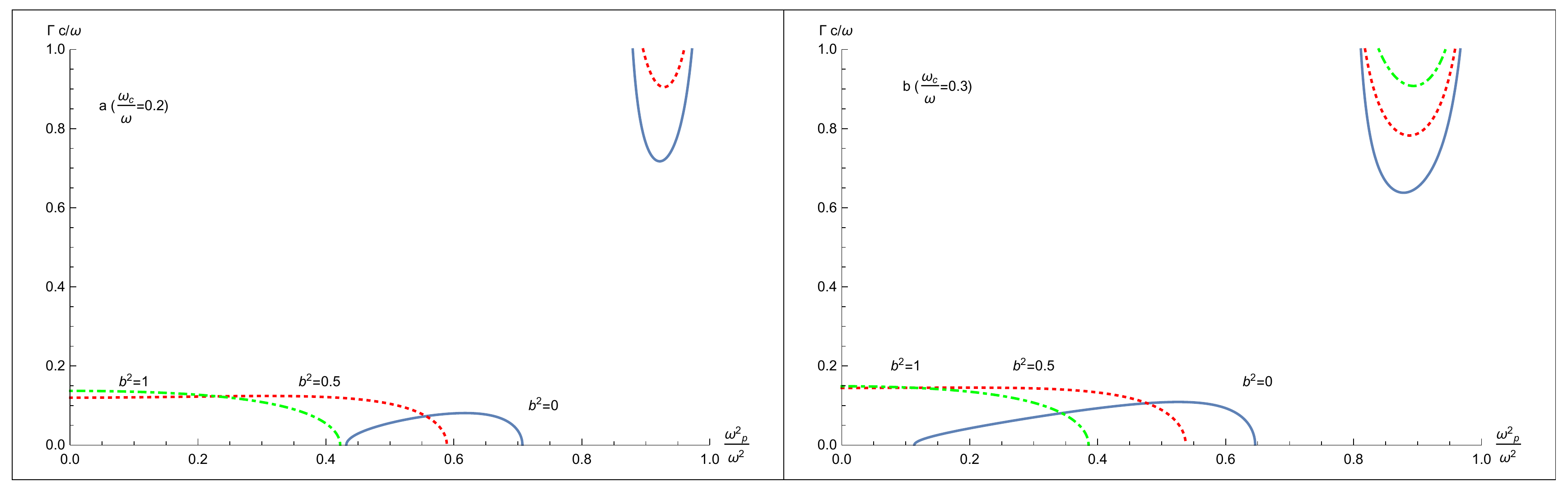}
\hspace{10mm}
\includegraphics[width=9cm]{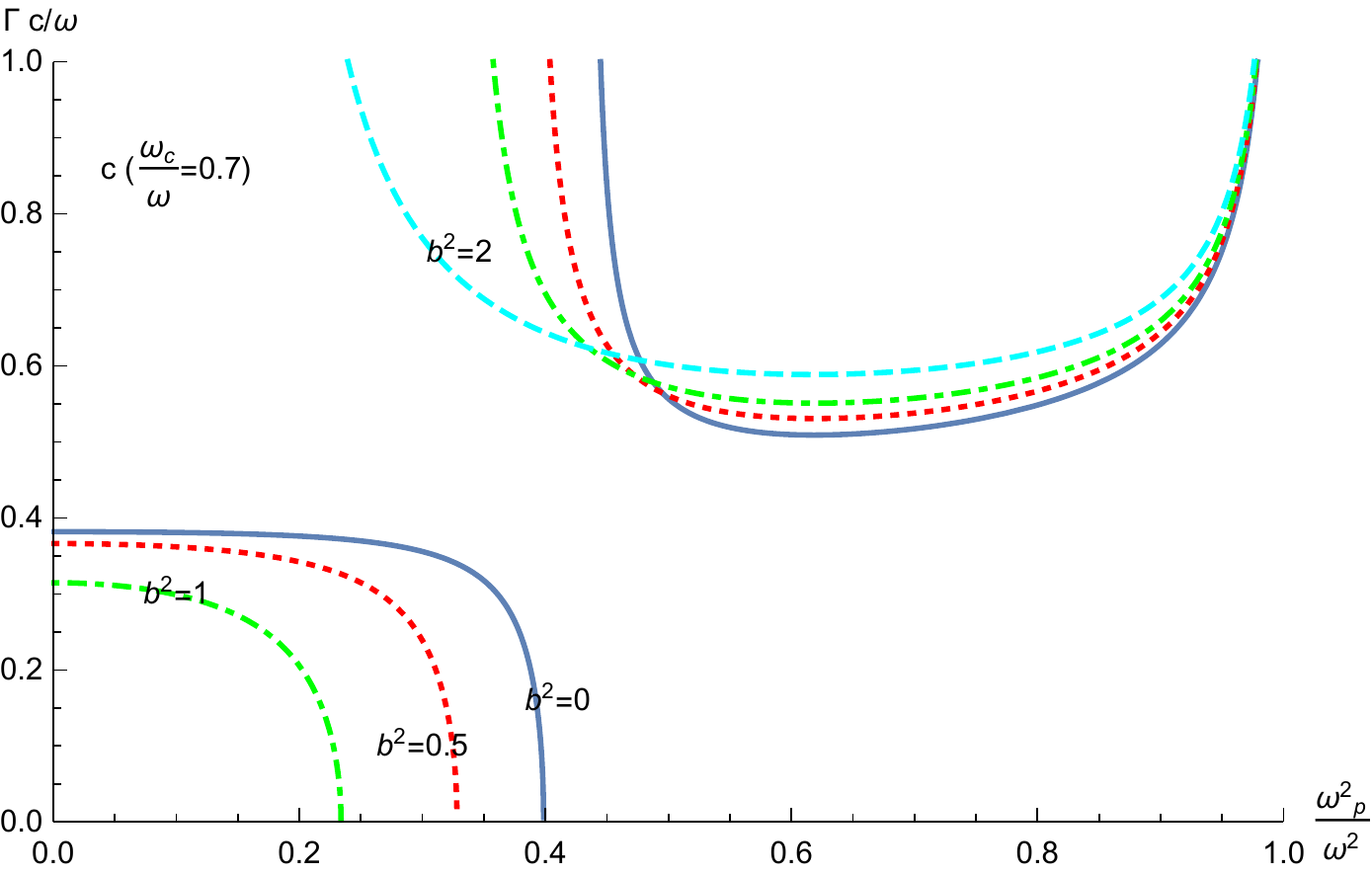}
\hspace{10mm}
\caption{ (Color online) normalized growth rate 
$\Gamma _{nor}(\Gamma c/\omega)$ versus 
$\frac{\omega^2_p}{\omega^2}$
for $\frac{\omega _c}{\omega}=0.2,0.3,0.7$, $a_0=1$, 
$d=\frac{1}{2}$, $Q_{ \perp }$=0.5 and $\gamma=\frac{1}{2}$.}
\label{fig4}
\end{figure}

\begin{figure}[!ht]
\centering
\includegraphics[width=12cm]{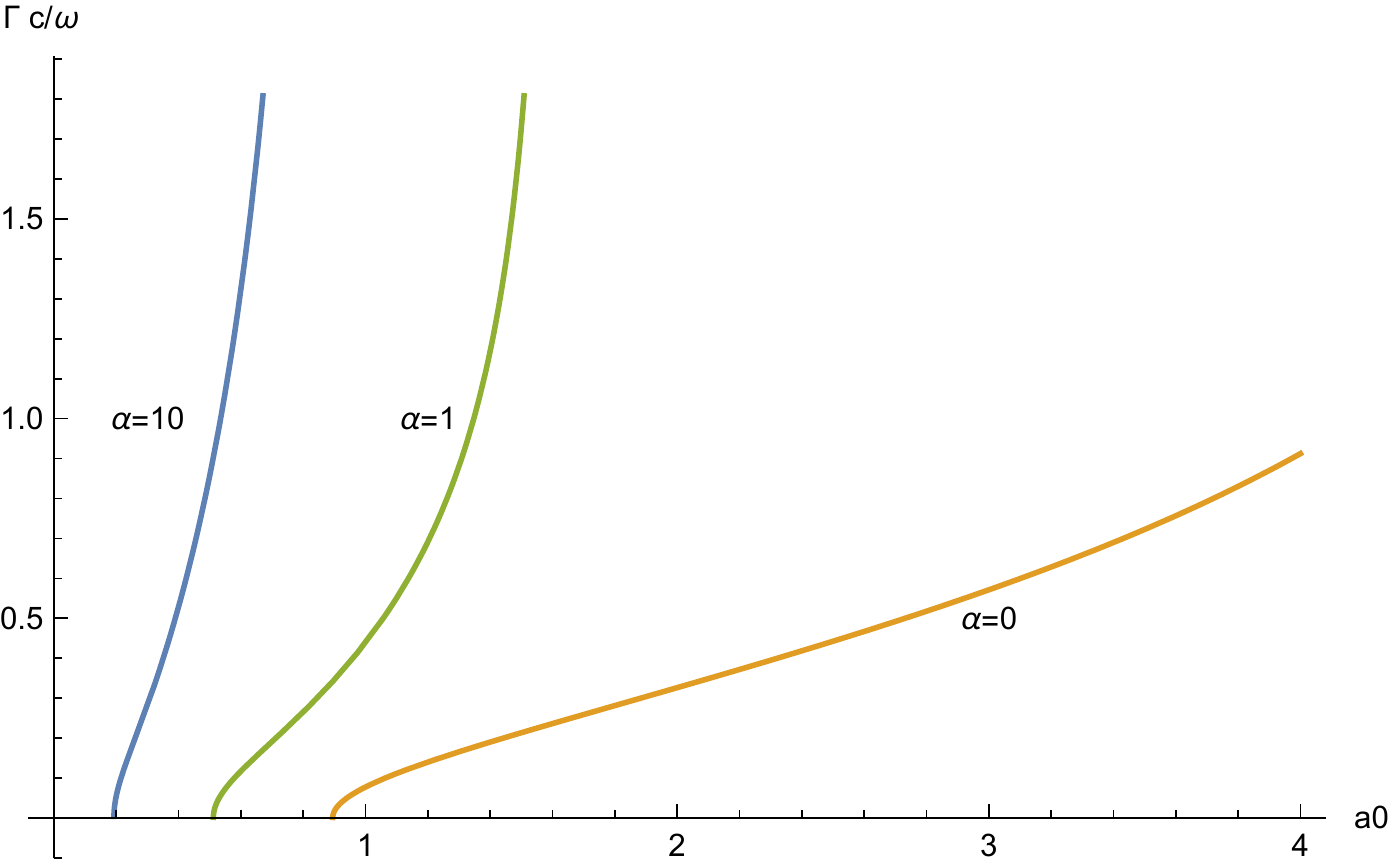}
\hspace{10mm}
\caption{ (Color online) normalized growth rate 
$\Gamma _{nor}(\Gamma c/\omega)$ versus
normalized laser field amplitude
$a_0=\frac{eA_{10}}{mc\omega}$
for $\frac{\omega _c}{\omega}=0.3$, $\frac{\omega_p^2}{\omega^2}=0.1$, 
$d=\frac{1}{2}$, $Q_{ \perp }$=0.5 and $\gamma=\frac{1}{2}$.}
\label{fig5}
\end{figure}

Since $\Gamma$ depends on the parameters involved, 
it cannot be determined without knowing the values of the parameters. 
The effects of each parameters can be seen 
by fixing the other parameters. The results are depicted in figures (\ref{fig1}-\ref{fig5}).
Fig. \ref{fig1} illustrates the normalized growth rate
$\Gamma$, $\Gamma \frac{c}{\omega}$,  
versus normalized transverse wave number 
$Q_{ \perp }$, $Q_L(q_\perp  c/\omega)$,
 for different values of Kerr parameters, $b^2$
and fixed other parameters. The results reveal that for a given $b^2$,
there is an optimum $Q_{ \perp }$, $Q_{ \perp }^{opt}$, which $\Gamma$
is maximum, $\Gamma_{max}$, at that point. This figure  shows
$\Gamma_{max}$ increases as $b^2$ increases.
Furthermore, the maximum is occurred in higher $Q_{ \perp }^{opt}$
for higher values of $b^2$.
The parameter $b^2$ is determined by non-linearity of the medium
and the intensity of laser pulse.
For example: for a typical peta-wat femtosecond laser pulse with
$\lambda$ =800nm and $|A_{10}|^2  \sim  10^{15}$, the value of the parameter
$n_2$ is obtained as order as $10^{-19}\frac{cm^2}{W}$ for gaseous media \cite{Ref22}.
This leads to the values 1-5 for the $b^2$ parameter.
Another important parameter is $\omega _c$
which can be easily adjusted by the external magnetic field, $B_0$.
Figs. \ref{fig2} and \ref{fig3} depict the mutual influence of the parameters
$\omega _c$ and $b^2$, for two values of
$Q_{ \perp }$ ($Q_{\perp}$=0.5 and $Q_{\perp}$= 0.75).
The results show that: for a specific values of $Q_{ \perp }$, there exist 
an optimum Kerr parameters, $b^{2}_{opt}$.
The loci of $b^{2}_{opt}$ can be adjusted by magnetic field via
$\omega _c$. $b_{opt}^{2}$ is smaller for larger
values of $\omega _c$.
The behavior of growth rate with respect to
$\omega _c$ and $b^2$, is heavily depends on the value of
$Q_{ \perp }$.
The results may reveal that the magnetic effects
could be competes by the non-linear effects.
Drawing normalized growth rate $\Gamma \frac{c}{\omega}$ versus
$\omega _c$ for different values of $b^2$, show that
normalized growth rate $\Gamma \frac{c}{\omega}$ is not a smooth function
of $\omega _c$, i.e. it is not a differentiable 
and single value function for all values of
$\omega _c$ e.g. for an interval
$\omega_c  \in [ \omega_{c1}, \omega_{c2}]$.
The value of $\omega_{c1}$ and $\omega_{c2}$
is depends on other parameters, particularly $b^2$
and $Q_{ \perp }$. The figures
resemblance the phase transition
phenomenon. The behavior of normalized growth rate
$\Gamma c/\omega$ before and after critical region is very different.
This strange behavior is also for variation of plasma density which
indicated by $\frac{\omega^2_p}{\omega^2}$.
This fact is depicted in Figs. \ref{fig4}a-\ref{fig4}c.
The result shows that the critical region and the behavior of
normalized growth rate
$\Gamma \frac{c}{\omega}$ before and after of this region, is heavily depends
on the parameters involved.
The figure \ref{fig5} illustrate the normalized growth rate versus
normalized laser field amplitude $( \sqrt{intensity})$
for some media.
To formalize the effect of medium Kerr non-linearity i.e.
$n_2$ we write $b^2= \alpha a^2_0$.
The numerical value of $\alpha$ is considered as
$\alpha=0$ (for linear medium),
$\alpha=1$ (for medium with weak non-linearity)
and $\alpha=10$ (for medium with high non-linearity).
The result show that the filamentation is not occurs 
in low intensity of laser for linear mediums.
Because each medium shows non-linearity when the power of the
driven laser exceeds a critical value, the filamentation is
observed only for $a_0 > a_{0c}$.
This figure yields $a_0  \sim 1$ leads to the value
$P_c\sim10^{15}\frac{W}{cm^2}$, which is in agreement 
with the result of other works. Also, this figure reveres 
that the growth rate of instability increases as
$a_0$ increase and there exist a saturation effect.
In fact, when filamentation is constructed, increasing the laser intensity,
increases the rate of filamentation growth.
But after an specific value of $a_0$ (intensity), the
filamentation growth rate reaches a constant value i.e. it
does not increase with the laser intensity. The physical region
of this phenomenon is may be related to the fact that after
the above mentioned laser intensity the filamentation
converts the absorbed energy of the laser to increase its temperature.
This fact ceases the growth of the instability.
Also this figure show that the intensity which the filamentation
stops its growth is lower for higher value of $\alpha$ i.e. higher values of Kerr coefficient.
\section{CONCLUSION}
The present study investigated the growth rate of filamentation
instability under the effect of non-linear polarization in magnetized plasma. The
results show that the non-linearity of medium  has
a significant influence on the formation rate of the filamentation. 
The growth rate of filamentation in non-linear medium 
can be controlled by external adjustable parameters such that
magnetic field, laser frequency and laser intensity and also
by varying internal parameters of the medium such that electron density
profile and Kerr coefficient. The results reveal that
there is an optimum range for the parameters involved. 
Also, for some specific and critical range of
parameters the growth rate of filamentation change
in undefined manner, reminds the phase transition phenomenon. 
Size of the critical region depends on the parameter involved specially,
$\omega _c$ and $\omega _p$. Also,
plotting the growth rate of the filamentation versus normalized laser
field amplitude, $a_0$ show that the growth rate of 
instability could be controlled by $a_0$ within an specific range.
For $a_0$ below this range there is $a_0$ filamentation and above this
region the filamentation growth rate becomes intensity-independent.
The Kerr-coefficient of the medium determines 
the start and final point of this region.
The results could be employed for the formation of
long-distance plasma channels and there for 
is applicable in the fields of interaction of the
intense laser pulses with rare gases.

\end{document}